%% file: main.tex
\documentclass{sig-alternate} 
\pdfoutput=1 
\usepackage{mathptmx} 

\newcommand{\ignore}[1]{}
\usepackage{fancyhdr}
\usepackage[normalem]{ulem}
\usepackage[hyphens]{url}
\usepackage{microtype}

\usepackage[bookmarks=true,breaklinks=true,letterpaper=true,colorlinks,linkcolor=black,citecolor=blue,urlcolor=black]{hyperref}

\usepackage{lipsum}
\usepackage{ragged2e}
\usepackage{xspace}
\usepackage{graphicx}
\usepackage{relsize}
\usepackage{color, listings}          
\usepackage{array,colortbl,xcolor}
\usepackage{caption}
\usepackage{subcaption}
\usepackage{framed}
\usepackage{multirow}
\usepackage{adjustbox}
\usepackage{paralist}
\usepackage{enumitem}
\setitemize{noitemsep,topsep=0pt,parsep=0pt,partopsep=0pt}

\newcommand\urmish[1]{\textcolor{green}{#1}}

\lstdefinestyle{customc}{
  belowcaptionskip=1\baselineskip,
  breaklines=true,
  xleftmargin=\parindent,
  xleftmargin=.2in,
  language=C,
  showstringspaces=false,
  basicstyle=\small,
  keywordstyle=\bfseries\ttfamily\color{blue},
  commentstyle=\itshape\color{black},
  identifierstyle=\ttfamily\color{black},
  stringstyle=\itshape\color{NavyBlue},
  keywords={if, for, continue, break, pragma, SXAPP, parallel},
  numbers=left
}

\title{A Static Analysis-based Cross-Architecture Performance Prediction Using Machine Learning
\vspace{-0.55in}
} 
\numberofauthors{3} 
\author{
Newsha Ardalani\thanks{The work done while student at UW-Madison.}\\
       \affaddr{Baidu Research}\\
       \email{newsha@baidu.com}
\alignauthor
Urmish Thakker\footnotemark[1]\\
       \affaddr{Arm ML Research}\\
       \email{urmish.thakker@arm.com}
\alignauthor 
Aws Albarghouthi\\
       \affaddr{UW Madison}\\
       \email{aws@cs.wisc.edu}
\alignauthor 
Karu Sankaralingam\\
       \affaddr{UW Madison}\\
       \email{karu@cs.wisc.edu}
}

\begin{document}
\maketitle
\pagestyle{plain}

\input{macros}

\input{abstract}
\input{intro}

\input{ml}

\input{features}

\input{analysis}
\input{related3}

\input{conc}


\bibliographystyle{ieeetr}

\clearpage
\input{dynamic}

\end{document}

%% file: macros.tex
\newcommand{\staticXAPP}[0]{XAPP-Static}
\newcommand{\fixme}[1]{\textbf{FIXME: #1}}
\newcommand{\todo}[1]{{\color{red}{\noindent\textsf{\textbf{{\scriptsize~TODO}}---#1}}}}

%% file: abstract.tex
\begin{abstract}
Porting code from CPU to GPU is costly and time-consuming; 
Unless much time is invested in development and optimization, 
it is not obvious, a priori, how much speed-up is achievable or how much room is left for improvement.
Knowing the potential speed-up a priori can be very useful: It can save hundreds of engineering hours, help programmers with prioritization and algorithm selection.

We aim to address this problem using machine learning in a supervised setting,
using solely the single-threaded source code of the program, without having to run or profile the code.
We propose a static analysis-based cross-architecture performance prediction framework (Static XAPP) which relies solely on program properties collected using static analysis of the CPU source code
and predicts whether the potential speed-up is above or below a given threshold.  
We offer preliminary results that show we can achieve 94\% accuracy in binary classification, in average, across different thresholds.
\end{abstract}

%% file: intro.tex
\section{Introduction}\label{sec:intro}
Porting code from CPU to GPU is a slow and tedious process; 
Not only the program needs to be restructured to extract maximum parallelism, 
data organization needs to be re-arranged too in order to benefit from different levels of GPU memory hierarchy.
A tool that can \textit{quickly} and
\textit{accurately} predict the speed-up 
could be extremely useful.
It can not only save programmers' time but also help with algorithm selection and prioritization.
From a programmers' standpoint, heuristic-based performance estimations are far from accurate;
GPU's architecture and programming paradigm are significantly different than CPU's 
and GPU programs' performance are very sensitive to events like branch divergence and memory divergence.
This work belongs to the broad category of performance prediction literature, which can be categorized as follows: 
\begin{itemize}
\item \textbf{Execution-based} techniques rely on dynamic binary
instrumentation to obtain program properties. Binary instrumentation slows the program execution by 10-1000$\times$ which can be very costly depending on the original program execution time. 
Collected features can be fed into a performance prediction
model, analytical or machine-learning, to obtain performance on a target machine~\cite{saavedra1996analysis,Hoste06performanceprediction,baldini2014predicting,ardalani}.
\item \textbf{Human-based} approaches like 
Roofline model~\cite{Williams:2009:RIV:1498765.1498785} and Boat-hull model~\cite{boat-hull1} avoid the 
overhead of binary instrumentation but relies on humans to estimate features, and thus can be imprecise and slow.
\item \textbf{IR-based} approach; we introduce this novel branch which relies merely on information available at the intermediate representation (IR) of a program, 
and thus avoid human involvement and slowdowns of binary instrumentation.
We make this insightful observation that program properties obtainable with simple static analysis
are sufficiently explanatory to predict cross-architecture performance.
This observation does not imply that the correlation between static program properties and speedup is by any means straightforward. In fact, we require sophisticated machine learning algorithms to discover this correlation.
\end{itemize}
We envision this tool to be useful/integrated in different scenarios including:
\begin{itemize}[leftmargin=0.1in]
    \item \textbf{Integrated Development Environments (IDEs)}; having an IDE environment where a developer can highlight a portion of the code to estimate the possible speed-up on a platform of choice can be highly useful. 
    \item \textbf{Device Placement Optimization}: Device placement optimization algorithms can benefit from an accurate prediction of speed-up when a particular algorithm is executed on a specific platform. An execution-based method can significantly slow-down such an algorithm and a human-based method will require continuous feedback. A tool like ours can help quickly and without intervention, to determine the possible speed-ups for various devices. 
\end{itemize}

%% file: ml.tex
\section{Methodology}\label{sec:ml}
We are operating within a small dataset regime as the size of our dataset is very small (156 datapoints). 
We briefly explain our machine learning approach, including the preparation phase, model construction phase, the details of the training and test sets, and the software/hardware platforms used in evaluation.


\textbf{Notations}
A $\emph{datapoint}$ is a pair of single-threaded CPU code and the associated GPU code. The CPU code is characterized in terms of its feature vector and the GPU code is used to measure the CPU-to-GPU speedup.
A $\emph{feature vector}$ is the set of program properties, outlined in Section~\ref{sec:feature}, estimated per CPU code and presented in the form of a binary vector.
%

\textbf{Preprocessing Steps}
Compared to dynamic binary instrumentation, static analysis can be orders of magnitude faster.
However, the estimated features can be less precise as they lack information about the dynamics of execution.
We trade-off precision for accuracy by discretizing the estimated feature values into two to three levels, using the \emph{equal frequency binning} algorithm.
We also discretize the output value (speedup) into two ranges, low and high;
from the developer's perspective, the decision to port a code to GPU rests more on the range of the speedup achievable (low or high) and less on the actual value of the speedup.
However, depending on the importance of the kernel, what considered as a high speedup range for one case might be low for another.
Therefore, we allow the user to denote the cutoff that breaks the speedup range into low and high.
We use the user-provided cutoffs to label each datapoint in our training set before model construction.


\textbf{Machine Learning Approach}
We employ the random forest (RF) algorithm to construct a speedup classifier.
Our RF model is an ensemble of 1000 decision trees, where each tree is constructed using a random subset of features and training datapoints.
We identify a set of 10 program properties that are sufficiently accurate using static analysis (see Section~\ref{sec:feature}).
Alternatively, this problem could have been formulated as an end-to-end deep learning problem, where the CPU source code could have been parsed through a recurrent or transformer model to implicitly discover the features and predict the performance on the target accelerator. 
We could not use this approach as we were operating in a small dataset regime.

\textbf{Dataset}
We collect our datapoints from the widely-known GPU benchmark suites, including Lonestar~\cite{kulkarni2009lonestar}, Rodinia~\cite{rodinia}, and NAS~\cite{bailey1991parallel,pilla}.
The codes available in these suites are mainly well-suited datapoints for GPU by design, and thus our dataset is highly biased.
To balance our dataset, we develop our own microbenchmarks and add some negative examples -- obviously ill-suited codes for GPUs -- to our dataset.
Collectively, this effort will give us $\sim80$ datapoints, which we refer to as \textit{core kernels}. 
In order to increase our dataset size further, we use a set of tricks prescribed by ~\cite{ardalani}; 
we manually develop alternate CPU and GPU
implementations by perturbing core kernels. For example, we add or subtract a piece of code that is well-suited or ill-suited for GPU, to both CPU and GPU implementations.

\textbf{Evaluation}
We use \emph{leave-one-out cross-validation} (LOOCV) to evaluate the accuracy of our technique,
which is widely-used for evaluation of small dataset problems.

%% file: features.tex
\section{Program Features}\label{sec:feature}
\newcommand{\mem}{\textsc{mem}\xspace}
\newcommand{\ctrl}{\textsc{ctrl}\xspace}
\newcommand{\arith}{\textsc{arith}\xspace}
\newcommand{\mul}{\textsc{mul}\xspace}
\newcommand{\divs}{\textsc{div}\xspace}
\newcommand{\scos}{\textsc{scos}\xspace}
\newcommand{\elog}{\textsc{elogf}\xspace}
\newcommand{\sqt}{\textsc{sqrt}\xspace}
\newcommand{\linv}{\textsc{linv}}
\newcommand{\pband}{\textsc{pband}}
\newcommand{\kbody}{\textsc{kbody}\xspace}
\newcommand{\coal}{\emph{coalesced}}
\newcommand{\diverge}{\emph{diverge}}

\newcommand{\ero}{f_I}
\newcommand{\ks}{\emph{ksize}}

\begin{table}
  \centering
\begin{tabular}{ll}
\hline
Statement type & Description\\
\hline
\mem & Memory loads and stores\\
\arith & All arithmetic operations\\
~~\mul $\subseteq$ \arith & FP multiplication \\
~~\divs $\subseteq$ \arith & FP division \\
~~\scos $\subseteq$ \arith & FP sine and cosine\\
~~\elog $\subseteq$ \arith & FP logarithm and exponential\\
~~\sqt $\subseteq$ \arith & FP square root\\

\ctrl & Conditional control statements\\
\hline
\end{tabular}
\caption{Program statements}
\vspace{-0.20in}
\label{tbl:stmt}
\end{table}

\begin{table*}
    \smaller
  \centering
\begin{tabular}{lllll}
\hline
\# & Feature & Formal definition & Relevance of feature for GPU performance\\
\hline

1 & Memory coalescing &  $ {\sum_{s \in \mem,  \coal(s)} \ero(s)}/ {\sum_{s \in \mem} \ero(s)} $& Captures whether memory accesses are coalesced\\
2 & Branch divergence &  $ {\sum_{s \in \ctrl, \diverge(s)} \ero(s)}/ {\sum_{s \in \ctrl} \ero(s)} $& Captures vulnerability to branch divergence\\
3 & Kernel size (\ks)& $\sum_{s \in P} f_I(s)$ & Captures whether kernel is embarrassingly-parallel\\
4 & Available parallelism &  $\ero(s)$ s.t. $s$ is inner-most loop in \pband & Captures
GPU resource utilization\\
  &&&\\

5 & Arithmetic intensity  & $\sum_{s \in \arith} \ero(s) / {\sum_{s \in \mem} \ero(s)} $ &
Captures ability to hide memory latency\\
6 & Multiplication intensity & ${\sum_{s\in \mul }\ero(s)} / {\ks}$  & Exploits GPU's abundant mul units\\
7 & Division intensity & ${\sum_{s\in \divs }\ero(s)} / {\ks}$  & Exploits GPU's abundant div units\\
8 & Sin/cos intensity & ${\sum_{s\in \scos}\ero(s)} / {\ks}$  & Exploits GPU hardware support for SFU\\
9 & Log/exp intensity & ${\sum_{s\in \elog }\ero(s)} /{\ks}$ & Exploits GPU hardware support for SFU\\
10 & Square root intensity & ${\sum_{s\in \sqt}\ero(s)} / {\ks}$  & Exploits GPU hardware support for SFU\\
\hline
\end{tabular}
\caption{Program features, their formal definition, and how they
impact GPU speedup}
\vspace{-0.18in}
\label{tbl:ftr}
\end{table*}

In traditional machine learning approach, we need to manually define the essential set of features required for characterizing
a desired output.
Here, we describe a generic CPU program model and
an associated static analysis framework that computes a number of
important program features for GPU speedup prediction.

\subsection{Program Model}

We will assume that we are given a sequential CPU program $P$
in a standard representation (e.g., LLVM's intermediate representation).
Program instructions are categorized as shown in Table~\ref{tbl:stmt}.
We will use \mem to denote the set of all memory
load and store instructions that appear in $P$.
Similarly, we will use \arith to denote arithmetic operations in $P$,
and \ctrl to denote conditional branches.

\subsection{Program Features and Static Extraction}

Assume for the moment that for a given a program $P$,
the developer has annotated
the region of the code---the loop or loops---they wish
to parallelize. We call this region the \emph{parallel band} (\pband).
We refer to the rest of the code enclosed within the \pband, as \emph{kernel body} (\kbody).
Figure~\ref{fig:ex} explains this with a simple example.
In this example, the outer for-loop is the parallel band -- as indicated by \textit{\#pragma parallel SXAPP} -- 
and the region enclosed within (line 3-11) is the kernel body.
While our features are statically determinable,
for the purposes of illustration, we will assume that
we are given an input $I$ of the program $P$.
Using $I$, we can characterize the number of
times an instruction $s$ is executed as a function of $I$,
which we call the \emph{expected occurrence frequency} of $s$
and denote by $f_I(s)$.
Note that this function, $f_I$, can only be discovered
dynamically. However, as we shall see in Appendix, our approach
is robust to the values of $f_I$ and we can elide
$f_I$ computation.

The set of
(numerical) features computed from $P$ is formally defined
and described in Table~\ref{tbl:ftr}.
In what follows, we provide a thorough exposition
of these features and the rationale behind choosing them.
We note that, while these features are numerical,
they will be later \emph{discretized} automatically
by our machine learning algorithms.

\begin{figure}
  \center
  \begin{lstlisting}[numbers=left]
#pragma parallel SXAPP (1048576) //parallel band
for (i=0; i < num_elements; i++) {
  key=i; j=0;
  if (key == tree[0]){
    found++; continue;
  }
  #pragma SXAPP(16)
  for(j=0; j<depth-1; j++) {
    j = (j*2) + 1 + (key>tree[j]);
    if (key == tree[j]) {
      found++; break;
} } }
  \end{lstlisting}
\vspace{-0.15in}
\caption{Example CPU code}
\label{fig:ex}
\vspace{-0.25in}
\end{figure}

\textbf{1. Memory coalescing} is a high-impact feature on GPU speedup;
it captures the possibility of global memory accesses to be coalesced.
A non-coalesced memory access can reduce the global memory bandwidth efficiency to as low as 1/32,
which negatively affects the speedup~\cite{mc}.
Specifically, this feature characterizes the percentage
of memory instructions in the \kbody that are considered \emph{coalesced}.
We weight each operation $s \in \mem$ by its occurrence frequency $\ero(s)$. 
Given a memory operation $s \in \mem$, we consider $\coal(s)$
to be true \emph{iff} one of the following holds: 
(1) Its memory index expression is loop-invariant with respect to all the loops
within the \pband. Intuitively, this means that all threads  access
the same memory location. 
(2) Its memory-index expression is loop-invariant with respect to all the loops
within the \pband, except the innermost one. The innermost-loop induction variable 
should appear with a $multiplier \leq 1 $ in the memory-index expression.
Intuitively, this means that consecutive threads are accessing to
consecutive or same memory locations.
In our running example in Figure~\ref{fig:ex}, there are two memory operations:
$\texttt{tree[0]}$ is coalescable, as the memory index is loop-invariant;
$\texttt{tree[j]}$ is considered non-coalescable, 
as the memory index $j$ depends on $key$ which depends on $i$, the induction variable of the loop in \pband.


\textbf{2. Branch divergence} Branch divergence is a measure of how effectively the parallel resources on GPU are being utilized.
Specifically, we characterize branch divergence as 
the percentage of conditional statements in the program that are considered \emph{diverging}.
We weigh each operation $s \in \ctrl$ by its  occurrence frequency $\ero(s)$. 
For a branch $s \in \ctrl$, we consider $\diverge(s)$ to be true
\emph{iff} at least one of the conditional expressions
in $s$ is not loop-invariant with respect to the parallel band loops.
Intuitively, this means that the branch condition may differ in
different threads, therefore can potentially diverge.

\textbf{3. Kernel size} The \emph{kernel size} (\ks) feature is
the number of instructions in the \kbody of the given program,
where each instruction is weighted by its occurrence frequency.
This is used as an indication of the dynamic number of instructions
to appear in the GPU kernel, and to enable computation of the
\emph{intensity} features described below.
Generally, when the kernel size is very large, it suggests that there is a loop with 
data dependency
across its iterations inside the \kbody, otherwise the loop should have moved into the \pband.
Therefore, the large kernel size indicates that the kernel is not embarrassingly-parallel.

\textbf{4. Available parallelism} The \emph{available parallelism} feature
is an approximation of the number of GPU threads.
Specifically, available parallelism is approximated
as the occurrence frequency of the inner-most
loop in the parallel-band.
In our running example, the parallel band is comprised
of a single loop (the outer-most one), and therefore
occurrence frequency of  that loop provides
an indication of the number of GPU threads.
Available parallelism indicates whether 
GPU resources are fully utilized.

\textbf{5-10. Instruction intensities} The lower part of Table~\ref{tbl:ftr} contains
features that measure whether the CPU code,
when ported to GPU, will exploit the strengths of GPUs.
For instance, the \emph{arithmetic intensity} feature is a measure of how well the arithmetic operations can hide  memory latency,
 and is defined as the ratio of the number of arithmetic operations to the number of memory operations.
 To estimate the number of memory operations/arithmetic operations statically, we weigh each operation $s$ by its  occurrence frequency $\ero(s)$.

Similarly, other features in this category,
measure of how effectively special function units on GPU are utilized.
For instance, the ratio of the number of single-precision floating-point sin/cos operations to the total number of instructions.

\subsection{Expected Occurrence Frequency}
The above feature extraction assumed the existence of a function $f_I$
that specifies the expected occurrence frequency of each program instruction.
While $f_I$ is not statically determinable,
we have empirically validated that our model is robust to
changes in $f_I$. 
Specifically, the expected occurrence frequency of an instruction
is a function of (1) loop-trip counts of loops enclosing
the instruction, and (2) the probability of taking branches
that lead execution to the instruction.
In Appendix~A, we show that our technique is robust to variation in
loop-trip count and branch probability and it can predict speedup with 91\% accuracy,
with no knowledge about the dynamic input, using a simple heuristic.

%% file: analysis.tex
\section{Results and Analysis}\label{sec:static-analysis}

In what follows, we first show the accuracy of our model for \emph{a} binary speedup classifier.
Next, we show our technique is robust across different cutoffs and platforms.
Finally, we analyze accuracy for a multiclass classifier.

\input{tables/thacc.tex}
\subsection{Model Accuracy}\label{subsec:accuracy}
Table~\ref{table:thacc} summarizes the accuracy results for a speedup classifier with the cutoff at 3.
We classify the speedup as low or high with 94\% accuracy.
The Positive Predictive Value (PPV) and Negative Predictive Value (NPV) are 93\% and 98\%, respectively.
The high NPV value suggests that our tool is very effective in saving programmers' time from porting a low-speedup application to GPU. 

\input{graph/stability.tex}
\subsection{Model Stability}~\label{subsec:stability}


To study the impact of cutoff choice on accuracy, we vary
the speedup cutoff values from $0$ to $100$ in steps of 1.
For each cutoff, we relabel our dataset and construct a new model, and measure its LOOCV accuracy.
Figure~\ref{fig:stability}(a) shows the prediction accuracy for different cutoffs on one GPU platform ( platform 1 in Table~\ref{table:hardware}).
As shown, our technique maintains minimum, average and maximum accuracy of 79\%, 86\% and 97\%, respectively.
Note here that the slight differences in accuracy across different cutoffs is partly due to changes in the number of datapoints within each interval.
Too many or too little datapoints in a bin can bias the model and hurt the generalization accuracy.
Figure~\ref{fig:stability}(b) shows similar results for another GPU platform (Platform 2 in Table~\ref{table:hardware}).
Since speedup distribution is different across different platforms, we observe different accuracy results for different speedup cutoffs.
Our technique maintains minimum, average and maximum accuracy of 76.5\%, 83\% and 89\%, respectively, on the second platform.
In conclusion, our technique is robust to variations in cutoffs and platforms.
\input{tables/platform.tex}

\input{graph/precision.tex}
\subsection{Multi-class classification}
We also study if we can predict speedup at a finer granularity, in other words, classifying the speedup in more than two bins.
Figure~\ref{fig:precision} represents the minimum, maximum and average prediction accuracy, 
as we increase the number of bins from $2$ to $5$.
The minimum, maximum and average accuracy are measured across different models constructed with different speedup cutoffs.
For instance, the second bar (3 intervals) represents the accuracy across
all models constructed with two speedup cutoffs, $(x_1,x_2)$, 
where $x_1$ varies from $1$ to $20$ in
steps of $1$ and $x_2$ varies from $x$ to $100$ in steps of $1$.
As expected, the model accuracy drops as the number of intervals (classes) increases.
This is expected as we get less datapoints in each interval.

%% file: tables/thacc.tex
\begin{table}[]
\centering
\label{my-label}
\begin{tabular}{lllll}
\cline{2-3}
\multicolumn{1}{l|}{}                                                                 & \multicolumn{1}{l|}{\textbf{\begin{tabular}[c]{@{}l@{}}Predicted:\\ Low\end{tabular}}} & \multicolumn{1}{l|}{\textbf{\begin{tabular}[c]{@{}l@{}}Predicted:\\ High\end{tabular}}} &                         &  \\ \cline{1-4}
\multicolumn{1}{|l|}{\textbf{\begin{tabular}[c]{@{}l@{}}Actual:\\ Low\end{tabular}}}  & \multicolumn{1}{l|}{TN: 61}                                                            & \multicolumn{1}{l|}{FP: 7}                                                              & \multicolumn{1}{l|}{68} &  \\ \cline{1-4}
\multicolumn{1}{|l|}{\textbf{\begin{tabular}[c]{@{}l@{}}Actual:\\ High\end{tabular}}} & \multicolumn{1}{l|}{FN:1}                                                              & \multicolumn{1}{l|}{TP: 87}                                                             & \multicolumn{1}{l|}{88} &  \\ \cline{1-4}
\multicolumn{1}{l|}{}                                                                 & \multicolumn{1}{l|}{62}                                                                & \multicolumn{1}{l|}{94}                                                                 &                         &  \\ \cline{2-3}
\multicolumn{5}{l}{}                                                                                                                                                                                                                                                                                  \\
\multicolumn{5}{l}{\begin{tabular}[c]{@{}l@{}}Accuracy = (TP+TN)/(TP+TN+FP+FN) = 94\%\\ Positive Predictive Value (PPV) = TP/(TP+FP) = 93\%\\ Negative Predictive Value (NPV) = TN/(TN+FN) = 98\%\end{tabular}}                                                                                      
\end{tabular}
\caption{Binary classifier accuracy (speedup cutpoint = 3)}~\label{table:thacc}
\vspace{-0.25in}
\end{table}

%% file: graph/stability.tex
\begin{figure}[t]
\centering
  \begin{subfigure}[b]{0.5\textwidth}
  \includegraphics[width=\linewidth]{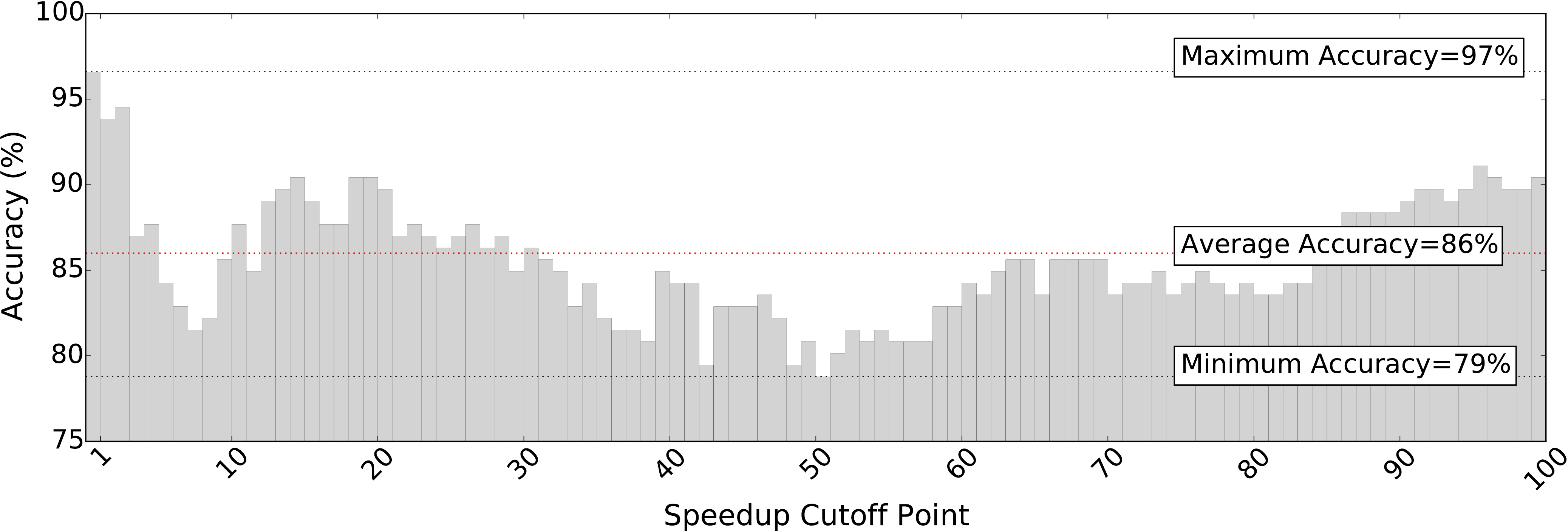}
  \vspace{-0.2in}
  \caption{Platform 1 (GTX750)}
  \label{fig:stability-platform1}
  \end{subfigure}
%
  \begin{subfigure}[b]{0.5\textwidth}
  \includegraphics[width=\linewidth]{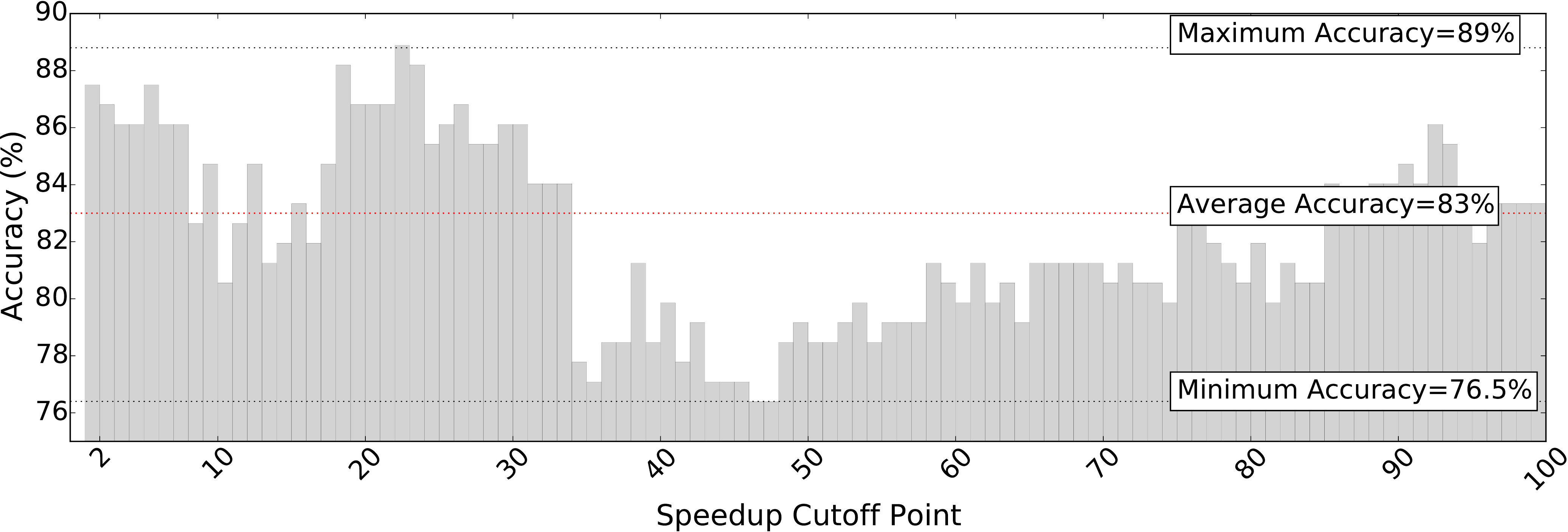}
  \vspace{-0.2in}
  \caption{Platform 2 (GTX660)}
  \label{fig:stability-platform2}
  \end{subfigure}

  \caption{Model Stability. The x-axis represents the cutoff point that divides the speedup range into low and high.
	Kernels with speedup $\leq$ x will be labeled as low (L) and kernels with speedup $>$ x will be labeled as high (H).
	The y-axis shows the cross-validation accuracy for a model that is constructed with a dataset labeled as such.} 
  \label{fig:stability}
  \vspace{-0.1in}
\end{figure}

%% file: tables/platform.tex
\begin{table}[t]
\small
\center
\begin{tabular}{llll}\hline
 & \textbf{Platform 1} & \textbf{Platform 2}\\ \hline
Micro\-architecture & Maxwell & Kepler \\ 
GPU model & GTX 750 & GTX 660 Ti \\ 
\# SMs & 7 & 14  \\
\# cores per SM & 192 & 192 \\
Core freq. & 1.32 GHz & 0.98 GHz \\
Memory freq. & 2.5 GHz &  3 GHz \\ \hline
CPU model: & \multicolumn{3}{l}{Intel Xeon Processor E3-1241 v3}\\ 
&\multicolumn{3}{l}{(8M Cache, 3.50 GHz)} \\ \hline
\end{tabular}
\caption{Hardware platforms pecifications.}\label{table:hardware}
\vspace{-0.25in}
\end{table}

%% file: graph/precision.tex
\begin{figure}
		\begin{center}
		\begin{framed}
		\includegraphics[width=\linewidth]{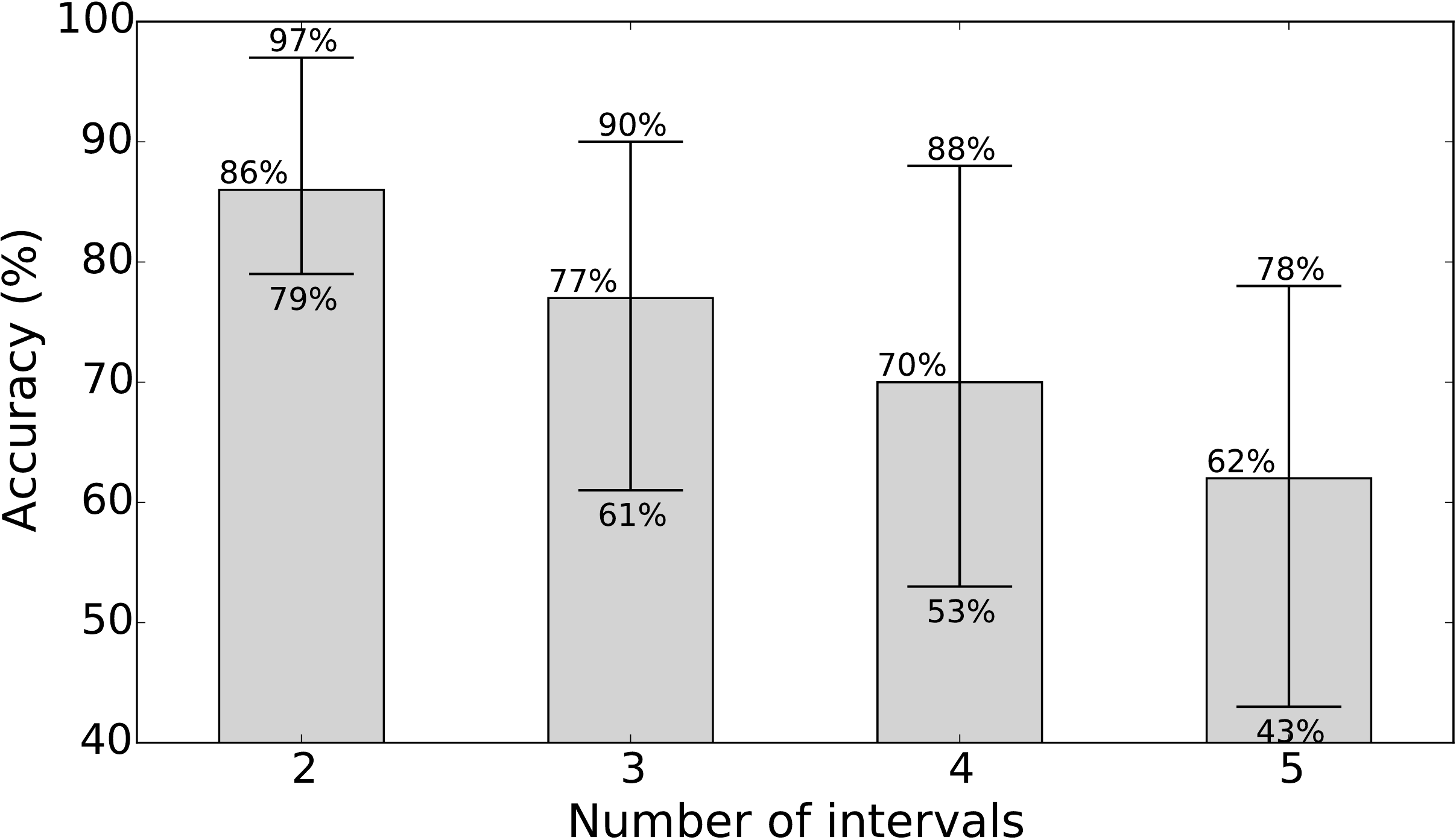}
		\newline
                \setlength{\tabcolsep}{.3em}
		\newline
		\footnotesize
		\renewcommand{\arraystretch}{1.5}
		\begin{tabular}{cc|cccc|}
		\cline{3-6}
		\multicolumn{1}{l}{}				 &   & \multicolumn{4}{c|}{Number of intervals} \\\cline{3-6}
		\multicolumn{1}{l}{}         &   & \multicolumn{1}{c}{2} & \multicolumn{1}{c}{3} & \multicolumn{1}{c}{4} & \multicolumn{1}{c|}{5} \\\hline
		\multicolumn{1}{|c|}{\multirow{5}{*}{\adjustbox{angle=90}{Cutoffs}}}& $x_1$ & (1,100,1)  	& (1,20,1) 	& (1,20,1) 		& (1,20,1)\\\cline{2-2}
		\multicolumn{1}{|c|}{} & $x_2$ & -      & ($x_1$,100,1)	& ($x_1$,100,5)		& ($x_1$,25,5)\\\cline{2-2}
		\multicolumn{1}{|c|}{} & $x_3$ & - 			& -             & ($x_2$,100,10)	& ($x_2$,50,10)\\\cline{2-2}
		\multicolumn{1}{|c|}{} & $x_4$ & -			& -			        & -               & ($x_3$,100,20)\\\hline
		\end{tabular}
		\end{framed}
		\end{center}
		  \caption{Multiclass classification accuracy. The Table below shows the range of cutoffs each bar is averaged across. 
			$(l,u,s)$ at row $i$ and column $j$ shows that cutoff $x_i$ for $j$ intervals sweeps between $l$ and $u$ in steps of $s$.}
		  \label{fig:precision}
		\vspace{-0.17in}
\end{figure}

%% file: related3.tex
\section{Related Work}\label{sec:rel}
The application of program analysis in cross-platform performance prediction has been previously explored, 
primarily in the context of design space exploration 
~\cite{yi2003statistically, vandierendonck2004many, ipek2006efficiently, wu2012inferred, Lee:2006:AER:1168857.1168881, 4147674}, finding the best CPU platform amongst 
many CPU microprocessors with different ISAs based on program similarity~\cite{saavedra1996analysis, johnperformance, Hoste06performanceprediction},
understanding performance bottlenecks of multicore architectures~\cite{Williams:2009:RIV:1498765.1498785},
and finding GPU acceleration based on CPU implementation~\cite{meng2011grophecy, meswani2013modeling, baldini2014predicting, ardalani}. 
As discussed in Section~\ref{sec:intro}, all of these studies are either execution-based, which introduces 10-100$\times$ slowdown, or human-based,
which is slow and imprecise.
Compiler community has explored techniques to automate GPU code generation from CPU code~\cite{schweitz2006r, ueng2008cuda, mikushin2012kernelgen, jablin2013automatic}.
However, their scope of applicability is limited to affine programs.
Hoshin et. al.~\cite{hoshino2013cuda} shows that
GPU codes generated OpenACC are 50\% slower than hand-optimized ones.
Static analysis has been previously used in the context of program optimization 
to predict the impact of an optimizion on performance~\cite{dubach2007fast}.
Many researchers have investigated GPU design space exploration 
and performance prediction
based on GPU program properties
~\cite{hongkimperf, 6189201,jia2013starchart,Baghsorkhi:2010:APM:1693453.1693470}.
However, these techniques require a GPU code to start with.

%% file: conc.tex
\section{Conclusion}\label{sec:conc}
In this paper we have developed a new speedup prediction technique
that relies only on the source code. It has
been believed that program properties needed for predicting GPU
speedup must necessarily be obtained from the dynamic execution of
the program.
Our paper makes a fundamental intellectual contribution in
demonstrating that statically determinable program properties are
sufficiently explanatory  for
developing a machine-learning based speedup predictor. 

%% file: dynamic.tex
\section*
{Appendix A. Impact of Input Dynamics on Speedup Range Prediction} \label{sec:dynamic}
The dynamic value of program properties defined in Section~\ref{sec:feature} is primarily controlled by two input-dependent factors: the number of iterations of each loop (also referred to as loop trip count) and the probability of each branch being taken (also referred to as branch taken probability).
We make this insightful observation that speedup is mostly robust to changes in the loop trip count and branch taken probability. Therefore, a simple heuristic such as the majority vote selection over
different predicted values of speedup for different values of loop trip counts and branch taken probabilities is a good enough estimator of the actual speedup range.
Specifically, we sweep each loop's trip count from 10 to 1000 and each branch taken probability from 0\% to 100\% and 
predict the speedup for each value of trip count and branch taken probability and get the majority vote over the predicted values. 
We show that this simple technique is surprisingly sufficient to capture the impact of program dynamics, 
and thus we can predict speedup with 91\% accuracy, with no knowledge about the dynamic input.
The reason the feature vectors, and consequently speedup range is robust to variations in dynamic variables can be summarized as follows: 
(1) The features are defined as ratios of two dynamic events, and usually numerator and denominator scale similarly when dynamic variable changes.
(2) They are discretized, which makes the discretized feature values robust to small changes in their actual value. 
(3) Cutpoints are close to the extreme ends, so the change in dynamic value of the feature usually keep it within the same region.
(4) Variations in the dynamic features that control performance, often affect CPU and GPU execution time 
in the same direction, and therefore speedup range remains unchanged.


\textbf{Sensitivity to Loop Trip Count}
\input{graph/sweepbranch.tex}
Figure~\ref{fig:sweepbranch} shows the distribution of speedup predictions while we vary the branch probability for each branch from 0\% to 100\% in steps of 25\%.
On the X-axis, we have all the kernels in our dataset that have at least one conditional branch within their kernel body.
Each stack bar represents the probability that a speedup prediction belongs to any of the following speedup ranges, $(0, 3)$ 
, $(3, 20]$, and $(20,\inf)$.
Specifically, we use our static-analysis tool to estimate the feature vector per a branch probability combination, 
and feed it into our speedup prediction model to get one speedup range prediction per a branch probability combination.
To give an example, $b+tree1-1-rd$, a kernel with three conditional branches, gets $5^3$ feature vectors and $5^3$ speedup range predictions.
\input{graph/sweeploop.tex}
Across all 125 predictions, the majority vote heuristics suggests that the speedup belongs to 3 to 20 range. 
To fact check this, we show the actual speedup value for each application by hatching the corresponding stack bar.
As shown, in all but three cases the actual speedup range matches with the majority prediction.
This implies that majority vote heuristic is sufficient to predict the speedup range with 91\% accuracy, 
with no knowledge about the branch probabilities.

\textbf{Sensitivity to Branch Taken Probabiltiy}
Figure~\ref{fig:sweeploop} shows the similar results for loop trip count.
On the X-axis, we have all the kernels in our dataset that have at least one loop within their kernel body.
We vary the values of trip-counts from 1 to 1000 in logarithmic steps and use majority vote prediction to predict the speedup. 
As shown, in all but two cases the actual speedup range matches with the majority vote prediction.
This indicates that majority vote heuristic is sufficient to predict the speedup range with 91\% accuracy, 
with no knowledge about the loop trip counts.

%% file: graph/sweepbranch.tex
\begin{figure}[t]
  \begin{center}
  \includegraphics[width=\linewidth]{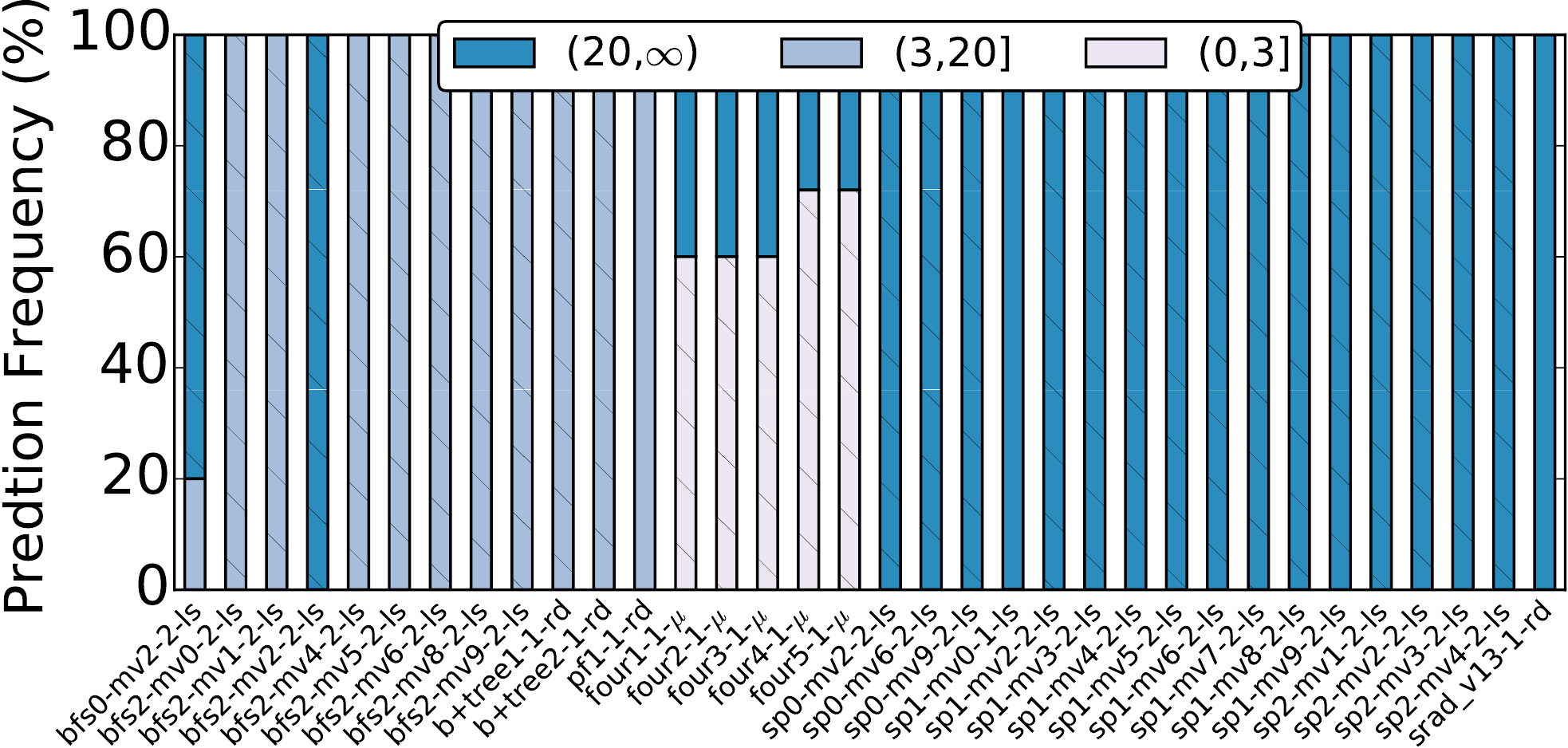}
  \end{center}
  \captionsetup{justification=centering}
  \vspace{-0.07in}
  \caption{Speedup prediction sensitivity to branch ratio.}
  \label{fig:sweepbranch}
  \vspace{-0.19in}
\end{figure}

%% file: graph/sweeploop.tex
\begin{figure}[!ht]
  \begin{center}
  \includegraphics[width=\linewidth]{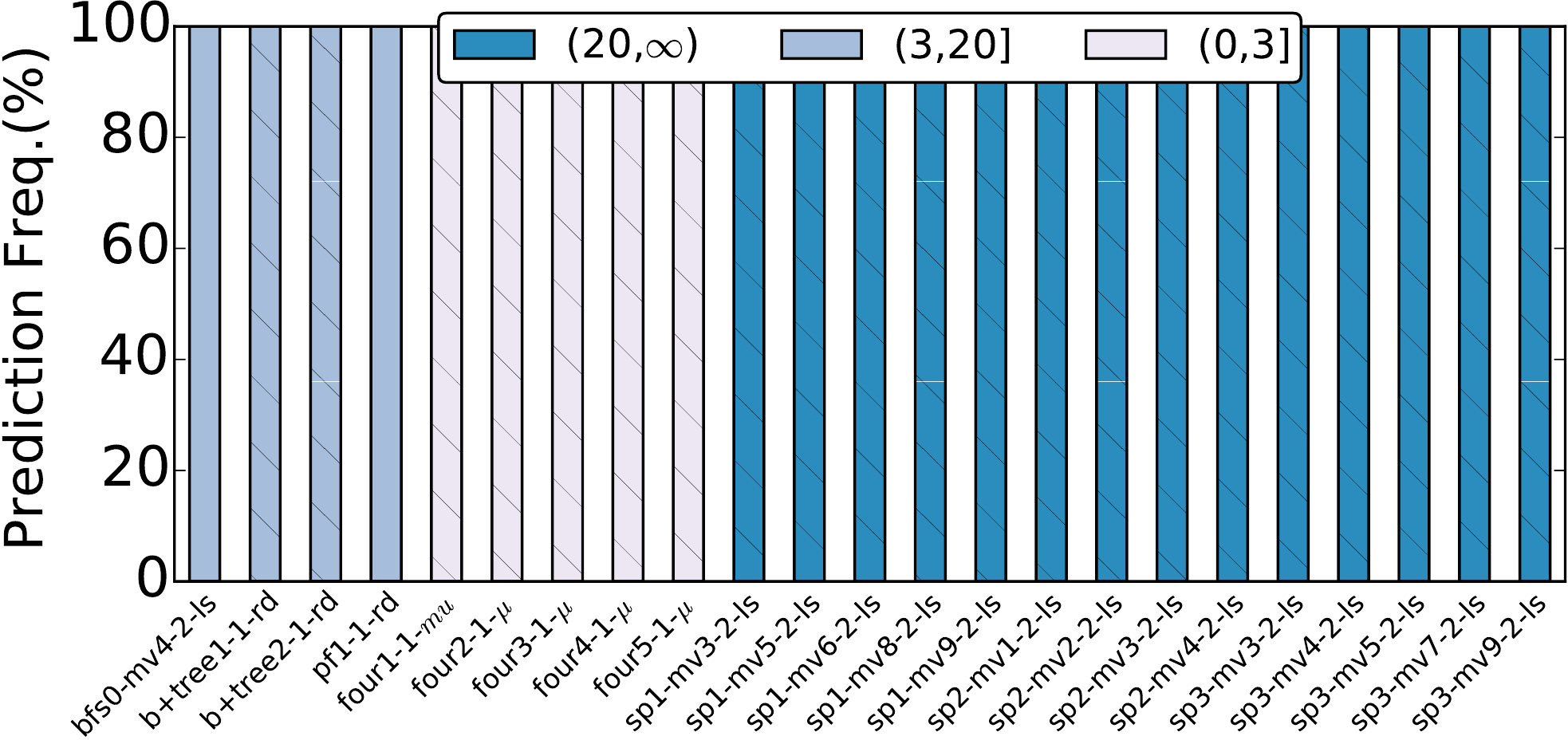}
  \end{center}
  \captionsetup{justification=centering}
  \caption{Speedup prediction sensitivity to loop trip count.}
  \label{fig:sweeploop}
\end{figure}

%% file: main.bbl
\begin{thebibliography}{10}

\bibitem{saavedra1996analysis}
R.~H. Saavedra and A.~J. Smith, ``Analysis of benchmark characteristics and
  benchmark performance prediction,'' {\em TOCS}, 1996.

\bibitem{Hoste06performanceprediction}
K.~Hoste, A.~Phansalkar, L.~Eeckhout, A.~Georges, L.~K. John, and K.~D.
  Bosschere, ``Performance prediction based on inherent program similarity,''
  in {\em PACT}, 2006.

\bibitem{baldini2014predicting}
I.~Baldini, S.~J. Fink, and E.~Altman, ``Predicting gpu performance from cpu
  runs using machine learning,'' in {\em SBAC-PAD '14}.

\bibitem{ardalani}
N.~Ardalani, C.~Lestourgeon, K.~Sankaralingam, and X.~Zhu, ``Cross-architecture
  performance prediction (xapp): Using cpu code to predict gpu performance,''
  in {\em MICRO}, ACM, 2015.

\bibitem{Williams:2009:RIV:1498765.1498785}
S.~Williams, A.~Waterman, and D.~Patterson, ``Roofline: an insightful visual
  performance model for multicore architectures,'' {\em Commun. ACM}, 2009.

\bibitem{boat-hull1}
C.~Nugteren and H.~Corporaal, ``The boat hull model: adapting the roofline
  model to enable performance prediction for parallel computing,'' in {\em
  PPOPP '12}, pp.~291--292, 2012.

\bibitem{kulkarni2009lonestar}
M.~Kulkarni, M.~Burtscher, C.~Cas{\c{c}}aval, and K.~Pingali, ``Lonestar: A
  suite of parallel irregular programs,'' in {\em ISPASS}, 2009.

\bibitem{rodinia}
S.~Che, M.~Boyer, J.~Meng, D.~Tarjan, J.~W. Sheaffer, S.-H. Lee, and
  K.~Skadron, ``Rodinia: A benchmark suite for heterogeneous computing,'' in
  {\em IISWC '09}.

\bibitem{bailey1991parallel}
D.~H. Bailey, E.~Barszcz, J.~T. Barton, D.~S. Browning, R.~L. Carter, L.~Dagum,
  R.~A. Fatoohi, P.~O. Frederickson, T.~A. Lasinski, R.~S. Schreiber, {\em
  et~al.}, ``The {NAS} parallel benchmarks,'' {\em IJHPCA}, 1991.

\bibitem{pilla}
L.~L. Pilla, ``{NAS Parallel Benchmarks CUDA} version.''
\newblock \url{http://hpcgpu.codeplex.com}.

\bibitem{mc}
``Cuda {Toolkit} {Documentation}.''
\newblock \url{http://docs.nvidia.com/cuda/cuda-c-best-practices-guide/}.

\bibitem{yi2003statistically}
J.~J. Yi, D.~J. Lilja, and D.~M. Hawkins, ``A statistically rigorous approach
  for improving simulation methodology,'' in {\em HPCA}, 2003.

\bibitem{vandierendonck2004many}
H.~Vandierendonck and K.~De~Bosschere, ``Many benchmarks stress the same
  bottlenecks,'' in {\em Workshop on Computer Architecture Evaluation Using
  Commercial Workloads}, 2004.

\bibitem{ipek2006efficiently}
E.~{\"I}pek, S.~A. McKee, R.~Caruana, B.~R. de~Supinski, and M.~Schulz, {\em
  Efficiently exploring architectural design spaces via predictive modeling}.
\newblock 2006.

\bibitem{wu2012inferred}
W.~Wu and B.~C. Lee, ``Inferred models for dynamic and sparse hardware-software
  spaces,'' in {\em MICRO}, 2012.

\bibitem{Lee:2006:AER:1168857.1168881}
B.~C. Lee and D.~M. Brooks, ``Accurate and efficient regression modeling for
  microarchitectural performance and power prediction,'' in {\em ASPLOS}, 2006.

\bibitem{4147674}
B.~Lee and D.~Brooks, ``Illustrative design space studies with
  microarchitectural regression models,'' in {\em HPCA}, 2007.

\bibitem{johnperformance}
A.~P. L.~K. John, ``Performance prediction using program similarity,''

\bibitem{meng2011grophecy}
J.~Meng, V.~Morozov, K.~Kumaran, V.~Vishwanath, and T.~Uram, ``Grophecy: Gpu
  performance projection from cpu code skeletons,'' in {\em SC}, 2011.

\bibitem{meswani2013modeling}
M.~R. Meswani, L.~Carrington, D.~Unat, A.~Snavely, S.~Baden, and S.~Poole,
  ``Modeling and predicting performance of high performance computing
  applications on hardware accelerators,'' {\em IJHPC}, 2013.

\bibitem{schweitz2006r}
E.~Schweitz, R.~Lethin, A.~Leung, and B.~Meister, ``R-stream: A parametric high
  level compiler,'' {\em HPEC}, 2006.

\bibitem{ueng2008cuda}
S.-Z. Ueng, M.~Lathara, S.~S. Baghsorkhi, and W.~H. Wen-mei, ``{CUDA-lite}:
  Reducing gpu programming complexity,'' in {\em LCPC}, 2008.

\bibitem{mikushin2012kernelgen}
D.~Mikushin and N.~Likhogrud, ``{KERNELGEN}--a toolchain for automatic
  gpu-centric applications porting,'' 2012.

\bibitem{jablin2013automatic}
T.~B. Jablin, {\em Automatic Parallelization for GPUs}.
\newblock PhD thesis, Princeton University, 2013.

\bibitem{hoshino2013cuda}
T.~Hoshino, N.~Maruyama, S.~Matsuoka, and R.~Takaki, ``Cuda vs openacc:
  Performance case studies with kernel benchmarks and a memory-bound cfd
  application,'' in {\em Cluster, Cloud and Grid Computing (CCGrid), 2013 13th
  IEEE/ACM International Symposium on}, pp.~136--143, IEEE, 2013.

\bibitem{dubach2007fast}
C.~Dubach, J.~Cavazos, B.~Franke, G.~Fursin, M.~F. O'Boyle, and O.~Temam,
  ``Fast compiler optimisation evaluation using code-feature based performance
  prediction,'' in {\em CF}, 2007.

\bibitem{hongkimperf}
S.~Hong and H.~Kim, ``An analytical model for a {GPU} architecture with
  memory-level and thread-level parallelism awareness,'' in {\em ISCA '09}.

\bibitem{6189201}
W.~Jia, K.~Shaw, and M.~Martonosi, ``Stargazer: Automated regression-based gpu
  design space exploration,'' in {\em ISPASS '12}.

\bibitem{jia2013starchart}
W.~Jia, K.~A. Shaw, and M.~Martonosi, ``Starchart: hardware and software
  optimization using recursive partitioning regression trees,'' in {\em PACT},
  2013.

\bibitem{Baghsorkhi:2010:APM:1693453.1693470}
S.~S. Baghsorkhi, M.~Delahaye, S.~J. Patel, W.~D. Gropp, and W.-m.~W. Hwu, ``An
  adaptive performance modeling tool for gpu architectures,'' in {\em PPoPP
  '10}.

\end{thebibliography}
